\documentclass[twocolumn,showpacs,preprintnumbers,amsmath,amssymb]{revtex4}
\usepackage{graphicx}
\usepackage{dcolumn}
\usepackage{bm}
\usepackage{here}

\begin{document}

\title{A Numerical Test of Pade Approximation for Some Functions 
with singularity}
\author{Hiroaki S. Yamada}
\email{hyamada[at]uranus.dti.ne.jp}
\affiliation{Yamada Physics Research Laboratory, 
Aoyama 5-7-14-205, Niigata 950-2002, Japan}
\author{Kensuke S. Ikeda}
\email{ahoo[at]ike-dyn.ritsumei.ac.jp}
\affiliation{Department of Physics, Ritsumeikan University
Noji-higashi 1-1-1, Kusatsu 525, Japan}

\date{\today} 
\begin{abstract} 
The aim of this study is to 
examine some numerical tests of Pade approximation for some typical 
functions with singularities such as 
simple pole, essential singularity, brunch cut and natural boundary. 
As pointed out by Baker, it was shown that the simple pole and 
the essential singularity 
can be specified by the poles of the Pade approximation. 
However, it have not necessarily been clear 
how the Pade approximation work for the functions 
with the branch cut or the natural boundary. 
In the present paper, it is shown that the poles and zeros of 
the Pade approximated functions are alternately lined 
along the branch cut if the test function has branch cut, 
and poles are also distributed around 
the natural boundary for some lacunary power series and random power series 
which rigorously has a natural boundary on the unit circle. 
On the other hand, spurious poles and zeros (Froissart doublets) 
due to numerical errors and/or external noise also appear 
around the unit circle in the Pade approximation. 
It is also shown that the residue calculus for the Pade 
approximated functions can be used to confirm the numerical 
accuracy of the Pade approximation and quasianalyticity 
of the random power series. 
\end{abstract}

 
\maketitle 
 
 
\def\ni{\noindent} 
\def\nn{\nonumber} 
\def\bH{\begin{Huge}} 
\def\eH{\end{Huge}} 
\def\bL{\begin{Large}} 
\def\eL{\end{Large}} 
\def\bl{\begin{large}} 
\def\el{\end{large}} 
\def\beq{\begin{eqnarray}} 
\def\eeq{\end{eqnarray}} 
 
\def\eps{\epsilon} 
\def\th{\theta} 
\def\del{\delta} 
\def\omg{\omega} 
 
\def\e{{\rm e}} 
\def\exp{{\rm exp}} 
\def\arg{{\rm arg}} 
\def\Im{{\rm Im}} 
\def\Re{{\rm Re}} 
 
\def\sup{\supset} 
\def\sub{\subset} 
\def\a{\cap} 
\def\u{\cup} 
\def\bks{\backslash} 
 
\def\ovl{\overline} 
\def\unl{\underline} 
 
\def\rar{\rightarrow} 
\def\Rar{\Rightarrow} 
\def\lar{\leftarrow} 
\def\Lar{\Leftarrow} 
\def\bar{\leftrightarrow} 
\def\Bar{\Leftrightarrow} 
 
\def\pr{\partial} 
 
\def\Bstar{\bL $\star$ \eL} 
 
\def\etath{\eta_{th}} 
\def\irrev{{\mathcal R}} 
\def\e{{\rm e}} 
\def\noise{n} 
\def\hatp{\hat{p}} 
\def\hatq{\hat{q}} 
\def\hatU{\hat{U}} 
 
\def\iset{\mathcal{I}} 
\def\fset{\mathcal{F}} 
\def\pr{\partial} 
\def\traj{\ell} 
\def\eps{\epsilon} 

 
\section{Introduction} 
Pade approximation was introduced in Mathematics 
and it has been
used in Physics for more than  40 years ago \cite{baker70,baker75,baker96}. 
In particular, there have been several important examples 
in Physics, to which Pade approximation was applied, such as
 summation of the divergent 
Rayleigh-Schrodinger perturbation series in 
scattering theory \cite{sasagawa81}, 
critical phenomena in statistical 
physics \cite{stanly71,yang52,nickel99,kubo85,macoy01}, 
denoising from noisy data of time-series, 
\cite{bessis96,stahl97,gilewicz97,bessis09,antonio09}, 
detection of singularity of phase space trajectories 
of Hamiltonian dynamical systems 
\cite{berrretti90,falcolini92,llave94b,berretti95,berretti01}, 
and so on. 
 
Mathematically, the Pade approximation is used 
to estimate analyticity of functions. 
Indeed, the Pade approximation is usually superior 
to the truncated Taylor expansions when 
the original function contains any singularity. 
Let us consider a simple example. 
The function $f(z)=\sqrt{\frac{1+2z}{1+z}}$ 
has brunch points at $z=-1$ and $z=-1/2$. 
The domain of convergence is $|z|<1/2$. 
Nevertheless, we can obtain an exact solution $\surd 2 =1.4142$ for 
$z\to \infty$ 
when we apply $[3|3]$ diagonal Pade approximation to the function. 
(See Sect. \ref{sect:test1} for more details of this example.) 
Although the mathematical validity of Pade approximation has not 
been exactly proved yet,  
The Pade approximation is practically very useful to continue a singular
function beyond the the domain of convergence.

Let us consider a critical phenomenon for Ising model 
as a simple example in statistical 
Physics \cite{stanly71,kubo85,nickel99}. 
We assume that at the critical point $u=u_c$ 
the exact magnetic susceptibility 
$\chi$ has a singularity as 
\beq 
\chi \sim (u-u_c)^{-\gamma }, 
\eeq 
where $u$ is a function of temperature and interactions and so on. 
In this case we sometimes use the logarithmic derivative of $\chi$ 
when we estimate the critical point $u=u_c$ 
and the critical exponent $\gamma$ as a pole-type singularity as, 
\beq 
\frac{d \log \chi}{d u}=\frac{ \chi^{'}}{\chi} = \frac{\gamma }{(u-u_c)}, 
\eeq 
where $\chi^{'}$ denotes the derivative 
with respect to the variable $u$. 
In the low temperature expansion for the magnetic susceptibility $\chi$ 
with some coefficients ${a_n}$, 
we assume that the approximated susceptibility $\chi_N$ 
and the logarithmic derivative 
are obtained as follows, 
\beq 
\chi \sim \chi_N = \sum_{n=0}^N a_n u^n, 
~~~ \frac{\chi^{'}}{\chi} \sim \sum_{n=0}^N b_n u^n. 
\eeq 
Then we can estimate the critical point $u=u_c$ and 
the critical exponent $\gamma$ 
applying the Pade approximation to the truncated expansion $\chi_N$. 
Note that the singularity on $|u|= u_c$ 
will be infinitely differentiable 
if the coefficients $a_n$ fall off sufficiently rapidly.

In general, analytic continuation over the singular point 
is possible along any other path in complex plane 
even if the function diverges at the singular point 
determining the radius of convergence, 
as seen in the above singularity of the Ising model. 
Therefore, the Pade approximation is useful 
to improve the convergence 
of the power series and approximate
 the exact solution.

Furthermore, the Pade approximation has been used to investigate 
convergence of Fourier series \cite{baker75,baker96} and 
the breakdown of KAM curves 
in complex plane for Hamiltonian map systems, 
which is described as the 
analytic domains of Lindstedt series for 
standard map 
\cite{berrretti90,falcolini92,llave94b,berretti95,berretti01}. 
 
In addition, we can see an interesting example concerning the 
Pade approximation in noisy data analysis 
\cite{bessis96,stahl97,gilewicz97,bessis09,antonio09}. 
The power series with finite random coefficient
``almost always'' has a natural boundary on the unit 
circle in the complex plane \cite{korner93,remmert10,breuer11,knill10,costin}. 
In a finite time-series, Froissart has shown that 
a natural boundary generated by the random time-series is approximated by 
doublets of poles and zeros (Froissart doublets) 
of the Pade approximated function 
surrounding the vicinity of the unit circle. 
Taking advantage of the characteristic, 
the Pade approximation has been used 
in order to remove the noise and extract 
the true poles associated with damping modes from 
the observed noisy time-series.

The main purpose of the present paper is to investigate 
whether the Pade approximation 
is numerically useful for detecting the singularity 
of some test functions. 
In particular, we examine the usefulness 
for the functions with a natural boundary 
such as a lacunary power series and a random power series 
\cite{baker75,berretti95,bessis09}.

The organization of the paper is the following: 
In Sect.\ref{sect:pade}, we give a brief explanation 
of the Pade approximation and 
some important reminders in the numerical calculation. 
In Sect.\ref{sect:pade_examples}, 
we present some numerical results of the Pade approximation for 
test functions with branch cut, essential singularity. We
also try to apply the Pade approximation to an entire function. 
In Sect.\ref{sect:nb1}, application of the Pade approximation 
to some lacunary series which is known to have a natural 
boundary is given.  In Sect.\ref{sect:nb2}, numerical results 
of the application to random power series with a natural boundary 
and some test functions with random noise are also shown. 
In Sect.\ref{sect:nb-random2}, we discuss about the residue calculus 
for the Pade approximated functions to confirm the numerical errors 
and quasianalyticity of the random power series. 
In the last section, we give summary and discussion.

In appendix \ref{app:recursion}, 
the general result for Fibonacci generator 
used in the Subsect.\ref{subsec:Fibo-generating} 
is given. 
In appendix \ref{app:random-polynomial}, 
some theorems concerning zeros of polynomials are given. 
In appendix \ref{app:lacunary-thorems}, 
some mathematical theorems for lacunary series which are useful in reading 
the main text, are summarized. 
In appendix \ref{app:bunshi}, exact Pade approximated function to 
some lacunary power series with a natural boundary are given. 
Residue analysis for quasianalytic functions of Carleman class is given 
in the appendix \ref{app:residue}. 
Furthermore, some theorems concerning the 
random power series are given in appendix \ref{app:noise1}.

\section{Pade approximation}
\label{sect:pade}
For a given function $f(z)$, a truncated Taylor expansion $f^{[N]}(z)$ 
of order $N$ about zero
 is given as, 
\begin{eqnarray}
     f(z) \sim f^{[N]}(z)  = \sum_{n=0}^{N} c_n z^n,
\label{eq:taylor1} 
\end{eqnarray}
where ${c_n}$ denotes the coefficients of the Taylor expansion.
Pade approximation is more accurate approximations for $f(z)$ up to order $O(z^{N})$
than the Taylor expansion.
The Pade approximation is a rational function, viz. a ratio of two polynomials, 
which agrees to the highest possible order $O(z^{N})$
with a truncated polynomial $f^{[N]}(z)$
 as follows.
\begin{eqnarray}
       f^{[N]}(z)      & = &  
 \frac{a_0+a_1z+a_2z^2+....a_Lz^L}{1+b_1z+b_2z^2+.......b_Mz^M}  \\
& \equiv &    \frac{P_L(z)}{Q_M(z)} \equiv f^{[L|M]}(z),
\label{eq:pade1}
\end{eqnarray}
where $P_L(z)$ is a polynomial of degree less than or equal to $L$ and 
$Q_M(z)$ is a polynomial of degree less than or equal to $M$.
Note that $b_0=1$ (normalized) here.
A unique approximation can be specified for all choice 
of $M$ and $L$ such that $N=L+M$ when it exists. 
The coefficients $\{a_n\},\{b_n\}$ can be obtained from the condition 
that the first $(L+M+1)$ terms vanish in the 
Taylor series. 
The difference between the Pade approximation and the original function 
satisfies a following equation,
\beq
 f(z) - f^{[L|M]}(z)  = O(z^{L+M+1}). 
\label{eq:pade2}
\eeq
In this paper, we sometimes use "$[L|M]$ Pade approximation" 
or "$[L|M]$ Pade approximated function" for $f^{[L|M]}(z)$.
Solving the problem in Eq.(\ref{eq:pade2}) is called 
a linear Teoplitze problem 
which is generally ill-conditioned.
We used full LU decomposition for the Teoplitze matrix in the problem
as well as iterative improvement in order to eliminate the ill-posed
problem \cite{press88}.
In addition, hereafter, we use the diagonal Pade approximation, i.e. $L=M$, 
of order $M \leq 65$ to estimate the singularity of the test functions
because of the convergence and limitation due to the round-off errors
and other source of numerical errors. 
Here, the singularity of the function $f(z)$ is 
approximated by configuration of the poles and zeros of the $[M|M]$
order diagonal Pade approximated function.
As mentioned in introduction, in general, 
Pade approximations are useful for representing 
unknown functions with possible poles.
The application of the diagonal Pade approximation 
is insured for the functions with isolated singular points and 
rational-type functions.
However, it is not fully-clarified that how the poles and zeros of 
the Pade approximated function describe 
 essential singularity, brunch cut and natural boundary.

Generally, the magnitude of the residues associated with 
the spurious poles are 
much smaller than those with the true poles, and they are close to 
machine precision. 
Very recently, Gonnet {\it et al.} suggested an efficient algorithm for 
the Pade approximations \cite{gonnet11,gonnet12}.
The algorithm detects and eliminates the spurious pole-zero pairs caused by 
the rounding errors by means of singular value decomposition for 
the  Teoplize matrix.

Before closing this section, we list up some important reminders
when we numerically apply the Pade approximation to unknown functions.

1. More accurate calculation becomes possible 
 by a scaling the expansion variable $z$
if there is a simple pole with large magnitude $\rho$($>>1$). 
That is, we should change the order of radius of convergence into $O(1)$ 
by the scaling the expansion variable as $z \to z/\rho$ in order to keep 
the numerical accuracy.
This procedure is effective when we apply the Pade approximation 
to exponentially 
decaying coefficients $\{a_n\}$ with fluctuation.

2. Poles (i.e. roots of $P_M(z)=0$ ) and
 zeros (i.e. roots of $Q_M(z)=0$) are 
sometime cancelled (zero-pole ghost pairs).
We can remove the effects of the ghost pairs 
and confirm the singularity of the functions by using 
 the residue analysis of the Pade approximated function.

3. Poles and zeros of the Pade approximation to the 
truncated random power series accumulate
 around the unit circle as Froissart doublets.
It is difficult to distinguish whether the poles of the 
Pade approximation originated from  
a natural boundary of the original function  or 
the natural boundary generated by the numerical error and/or noise.
Therefore, the numerical accuracy will be important 
to  determine the coefficients of the Pade approximation.

4. In general, the denominators of 
the diagonal Pade approximated functions to the lacunary power series 
and the random power series become lacunary 
and random polynomials, respectively. 
Accordingly, the distribution of the poles and zeros
of the approximated functions 
are similar to the distribution of the zeros corresponding 
to the original lacunary power polynomial 
and random power polynomial.  
In particular, it is well-known that the zeros of the random polynomials 
uniformly distribute about the unit circle. 

\section{Examples of Pade Approximation for Some Functions}
\label{sect:pade_examples}
In this section we investigate the configuration of the poles and zeros 
of the Pade approximation to some test functions with singularity.

\subsection{comparison of Pade approximation with Taylor expansion}
First, we try the Pade approximation to the following test function $f_1(z)$  
with a brunch point at $z=-1$, 
\beq
 f_{1}(z) &=& \frac{\log(1+z)}{z}. 
\eeq
The truncated Taylor expansion of the order $N=4$ 
around $z=0$ is
\beq
 f_{1}^{[4]}(z) &=& 1-\frac{1}{2}z+\frac{1}{3}z^2
-\frac{1}{4}z^3+\frac{1}{5}z^4. 
\eeq 
The $[2|2]$ Pade approximation is given as, 
\beq
 f_{1}^{[2|2]}(z) &=& \frac{1+\frac{7}{10}z
+\frac{1}{30}z^2}{1+\frac{6}{5}z+\frac{3}{10}z^2}. 
\eeq 

\begin{figure}[htbp] 
\begin{center} 
\includegraphics[width=6.5cm]{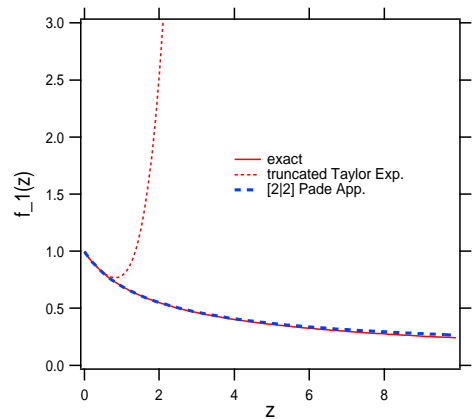}
\caption{
\label{fig:fig1}
(Color online)
The Pade approximated function $f_1^{[2|2]}(z)$, 
 the truncated Taylor series $f_1^{[4]}(z)$
and the original test function $f_1(z)$.
}
\end{center}
\end{figure}

Figure \ref{fig:fig1} shows the approximated functions 
the truncated Taylor series $f_1^{[4]}(z)$
and the original test function $f_1(z)$.
The approximated function $f_{1}^{[4]}(z)$ by the truncated Taylor expansion 
converges only within $|z|<1$ and 
deviates from the exact function $f_{1}(z)$ for $|z| > 1$.
On the other hand, 
it follows that the Pade approximated function $f_{1}^{[2|2]}(z)$ 
 well-approximates the original function $f_1(z)$
with very high precession even beyond the radius of convergence 
up to $Re z =x \sim 10$.
Therefore, it is found that the divergent power series expansion (Taylor expansion)
does still contain information 
about the original function outside the convergence radius, 
and rearranging the coefficients of the expansion into 
the Pade approximation recovers the information.
As a result, the conversion from the Taylor-form to the Pade-form usually accelerates 
the convergence and often allows good 
accuracy even outside the radius of convergence of the power series.

\subsection{Pade approximation for Fibonacci generating function}
\label{subsec:Fibo-generating}
Let us consider the Fibonacci generating function $f_F(z)$, 
where Fibonacci sequence $\{ F_n \}$ is encoded in the power series 
as the coefficients. 
The Fibonacci sequence $\{ F_n \}$ is 
given by the following recursion relation:
\beq
  F_n=F_{n-1}+F_{n-2} (n\geq 2),  
\eeq
where $F_0=0$ and $F_1=1$. 
Then the Fibonacci generating function $f_F(z)$ becomes 
\beq
 f_{F}(z) &=& \sum_{n=0}^{\infty} F_n z^n \\
       &=& \frac{z}{1-z-z^2}  \\ 
       &=& \frac{1}{\sqrt{5}} \frac{1}{1-\phi^+z}
-\frac{1}{\sqrt{5}} \frac{1}{1-\phi^-z}, 
\eeq
where $\phi^+ \equiv \frac{1+\sqrt 5}{2}(=1.61803.....)$ and 
$\phi^- \equiv \frac{1-\sqrt 5}{2}(=-0.61803...)$.
The generating function has poles at $z=\phi^+$ and $z=\phi^-$.
Generating functions by more general recursion relation is given in 
appendix \ref{app:recursion}.

 It should be noted that the diagonal Pade approximation 
$f_F^{[\frac{N}{2}|\frac{N}{2}]}(z)$ 
for 
the truncated Fibonacci generation function  
$f_F^{[N]}(z)=\sum_{n=0}^{N} F_n z^n$
has the form
\beq 
 f_F^{[\frac{N}{2}|\frac{N}{2}]}(z) = \frac{z}{1-z-z^2}  + O(z^{N+1}), 
\eeq 
for even number $N \geq 2$.
This means that the diagonal Pade approximation can detect the 
exact poles of the generating function irrespective of the order.

\subsection{Examples of some test functions with 
pole, brunch cut and essential singularity}
\label{sect:test1}
In this subsection, we use some test functions in applying of the Pade 
approximation.
\beq
f_2(z) &=& e^{-z},   \\
f_3(z) &=&  \sqrt{\frac{1+2z}{1+z}} , \\
f_4(z) &=& e^{-z/(1+z)}, \\
f_5(z) &=&  \tan z^4. 
\eeq
Here, $f_2(z)$ has no singularity for $|z|<\infty$, 
$f_3(z)$  has a brunch cut along a line on $[-1,-1/2]$, and 
$f_4(z)$  has an essential singularity at $z=-1$.
$f_5(z)$ has eight poles at points on the unit circle 
$z=\exp\{i\pi \frac{m}{4} \} (m=0,1,2,...,7)$.

\begin{figure}[htbp] 
\begin{center} 
\includegraphics[width=7cm]{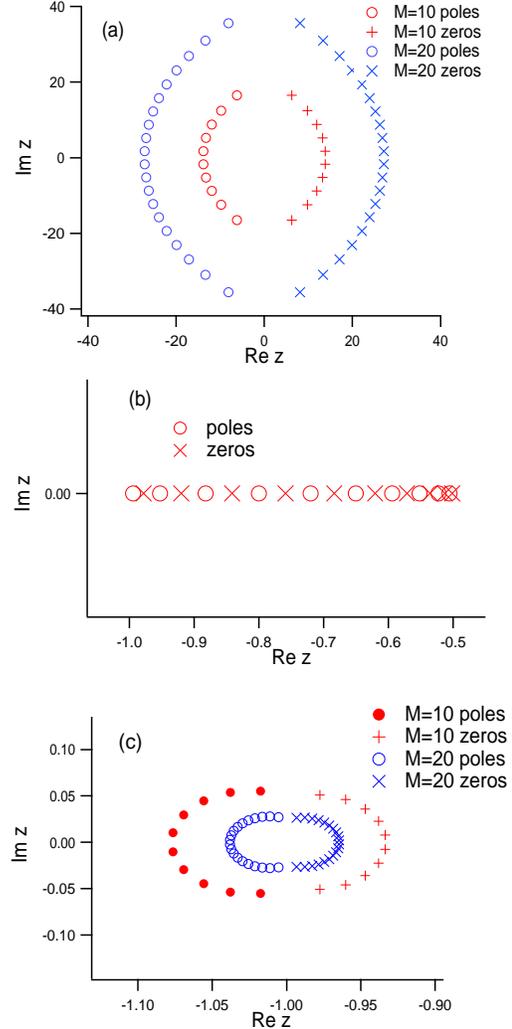}
\caption{\label{fig:fig2}(Color online)
Distributions of poles ($\bullet,\bigcirc$) and zeros ($+,\times $) 
of the $[M|M]$ diagonal 
Pade approximated functions 
 to some test functions:(a)$f_2(z)$, (b)$f_3(z)$, (c)$f_4(z)$.
}
\end{center} 
\end{figure}

First, let us apply Pade approximation to $f_2(z)$.
In this case the explicit form of the Pade approximated function
can be obtained in the following form,
\beq
  P_M(z) &=& \sum_{k=0}^{M} \frac{(2M-k)!M!}{(2M)!k!(M-k)!} (-z)^k, \\
  Q_M(z) &=& \sum_{k=0}^{M} \frac{(2M-k)!M!}{(2M)!k!(M-k)!} z^k.
\eeq
Note that the coefficients of the numerator $P_M(z)$
 have always alternatively sign,  
 and the zeros and poles of Pade approximated 
function are symmetrical to the imaginary axis with each other
because $P_M(z)=Q_M(-z)$.
Figure \ref{fig:fig2}(a) shows the numerical results 
in the complex $z-$plane.
All poles are on the left-half plane $Re z >0$, and 
all zeros are on the right-half plane $Re z <0$.
 The poles and zeros of the Pade approximated functions 
for the regular function $f_2(z)$ 
 go infinity and disappear as $M \to \infty$
because the function $f_2(z)$ is an entire function.

\begin{figure}[htbp]
\begin{center}
\includegraphics[width=8cm]{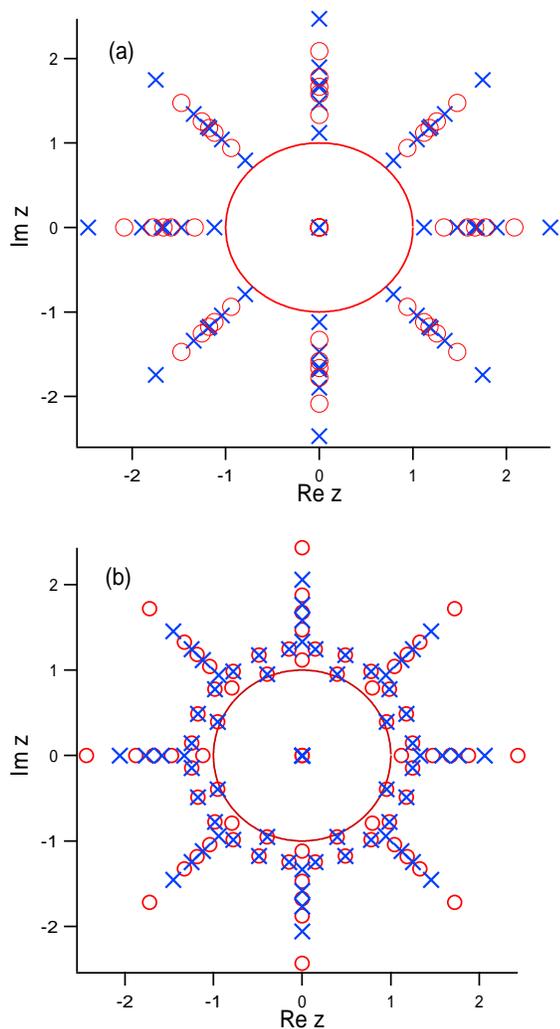}
\caption{
\label{fig:tan}(Color online)
Distributions of poles ($\bigcirc$) and zeros ($\times $) 
of the $[M|M]$ diagonal 
Pade approximated functions to the test function $f_5(z)$.
(a)The $[50|50]$ Pade approximation. (b)The $[75|75]$ Pade approximation.
The unit circle is drawn to guide the eye.
}
\end{center}
\end{figure}

In Fig.\ref{fig:fig2}(b), distribution of 
the zeros and poles of Pade approximated function to $f_3(z)$ are shown.
The poles and zeros make a line alternately 
between the two branch points of the $f_3(z)$; $z=-1$ and $z=-1/2$.

Figure \ref{fig:fig2}(c) shows the distribution of the zeros and poles of 
the Pade approximation to $f_4(z)$.
Pade approximation clusters the poles and zeros 
at the singular point for $f_4(z)$.
As $M \to \infty$ the poles and zeros 
approach the singular point $z=-1$ that reflects 
the essential singularity of the 
original function $f_4(z)$.

Figure \ref{fig:tan}(a) shows the poles and zeros of 
the Pade approximated function
to the test function $f_5(z)$.
The poles and zeros are alternatively distributed 
in eight-direction from the origin.
It seems that the distance between the poles and/or 
zeros on the same line becomes
small as they approach the location of the true poles. 
Some "spurious poles" appear around the unit circle 
as increase of the order of the 
Pade approximation, as seen in Fig.\ref{fig:tan}(b), 
which are irrelevant poles due to 
insufficient numerical accuracy.
It is found that the numerical accuracy of 
the Pade approximation fails for the 
higher order of the Pade approximation.
We discuss about the spurious poles 
in Sect.\ref{sect:nb1} and \ref{sect:nb2} again.

\section{natural boundary of lacunary power series}
\label{sect:nb1}
In this section we examine the applicability of the Pade approximation
to investigating the analyticity of some well-known
test functions with a natural boundary on the unit circle $|z|=1$.
This will provide a preliminary information about what
occurs in the Pade approximated functions with a natural boundary. 
Indeed, we do know only a very few numerical examples
which has a natural boundary and allows an exact diagonal Pade approximation.

There are following famous lacunary power series 
with a natural boundary on the unit circle $|z|=1$,
$f_{Jac}(z)=\sum_{n=0}^{\infty}  z^{2^n}$,
$f_{Wie}(z)=\sum_{n=0}^{\infty}  z^{n!}$,
$f_{Kro}(z)=\sum_{n=0}^{\infty}  z^{n^2}$,
where the $f_{Jac}(z)$, $f_{Wie}(z)$  and 
$f_{Kro}(z)$ are called after 
Jacobi, Weierstrass and Kronecker. 
Some theorems for the 
lacunary series with a natural boundary 
are given in appendix \ref{app:lacunary-thorems} 
\cite{korner93,remmert10,breuer11}.

Here, we use polar-form $f_r(\theta)$ for the function $f(z)$ by 
changing the variable, i.e. $z=re^{i\theta}$, 
 in order to simply display the functions as; 
\beq
 f_r(\theta) = f(z=re^{i\theta})=\sum_{n=0}^{\infty} c_n (re^{i\theta})^n.
\eeq
Then, note that the modulus $r$ works as a convergence factor of the series
because it well converges for $r<1$.
Typically we take $r=1$ on the unit circle or $r=0.98$ inside the circle
in the following numerical calculations.

\subsection{Example 1: Jacobi lacunary series}
\begin{figure}[htbp]
\begin{center}
\includegraphics[width=8cm]{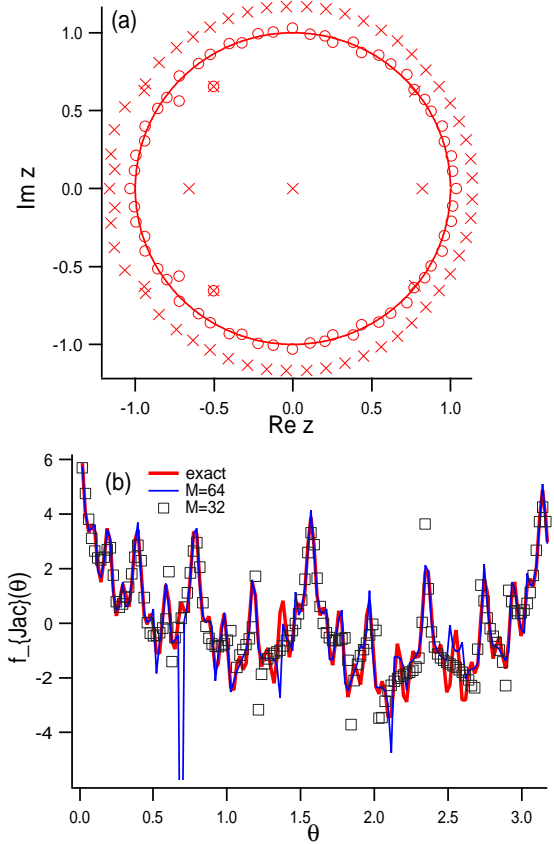}
  \caption{
\label{fig:fig3}(Color online)
(a)Distribution of poles($\bigcirc$) and zeros($\times$) 
of the $[64|64]$ Pade approximation  
to the test function $f_{Jac}(z)$ with a natural boundary on $|z|=1$.
The unit circle is drawn to guide the eye.
(b)The Pade approximated functions  $f_{Jac}^{[32|32]}(\theta)$,
 $f_{Jac}^{[64|64]}(\theta)$ and the exact function $f_{Jac}(\theta)$ 
in the polar-form with $r=1.0$. (After Ref.\cite{yamada13}.)
}
\end{center}
\end{figure}

We try to apply Pade approximation 
to the function $f_{Jac}(z)$ with a natural boundary on 
the unit circle $|z|=1$. 
The Pade approximated function exactly has the following form, 
\beq
f^{[2^N]}_{Jac}(z) & \sim & f_{Jac}^{[2^{N-1}|2^{N-1}]}(z)  \\
& = &  \frac{A^{N}_{Jac}(z)}{1+\sum_{k=0}^{N-2} z^{2^{k}} - z^{2^{N-1}}},
\label{eq:jac}
\eeq
where the explicit form of the numerator $A^{N}_{Jac}(z)$ 
is given in appendix \ref{app:bunshi}.
Accordingly, the poles of the $[2^{N-1}|2^{N-1}]$ Pade approximated function is 
given by roots of 
the polynomial, 
\beq
1+\sum_{k=0}^{N-2} z^{2^{k}} - z^{2^{N-1}} =0. 
\label{eq:jac2}
\eeq 
This is also just a lacunary polynomial.
In Fig.\ref{fig:fig3} the numerical result of 
the Pade approximation for $f_{Jac}(z)$ is shown.
The poles and zeros are plotted for the $[64|64]$ 
Pade approximation in Fig.\ref{fig:fig3}(a).
Inside the circle $|z|=1$ some cancellations of the ghost pairs appear.
The poles and zeros  accumulate around $|z|=1$ 
as increase of order of the Pade approximation.
In the case of the $M=64$, the poles accumulate around $|z|=1$  
with making the zero-pole pairings.
Figure \ref{fig:fig3}(b) shows the Pade approximated functions 
in the polar-form with $r=1$. 
It well approximate the original function $f_{Jac}(z)$ 
when the order of the Pade approximation
increases.

It is also shown that the complex zeros of 
the polynomial (\ref{eq:jac2}) cluster near 
unit circle $|z|=1$ and distribute uniformly on the circle
as $F_N \to \infty$ by Erdos-Turan type theorem given
in appendix \ref{app:lacunary-thorems} and \ref{app:random-polynomial}
\cite{kac43,erdos50,amoroso96,odlyzko93,simon04a,simon04b,peres05,simon10}.

\subsection{Example 2: Fibonacci lacunary series}
As a second example, we would like to apply Pade approximation
to the following lacunary series 
\beq
f_{Fib}(z) = \sum_{n=0}^{\infty} z^{F_n},  
\eeq
where $F_n$ is $n$th Fibonacci number.
 This function also has a natural boundary on $|z|=1$.
The Pade approximated function exactly has following form, 
\beq
f^{[F_N]}_{Fib}(z) & \sim & f^{[\frac{F_N}{2}|\frac{F_N}{2}]}_{Fib}(z) \\
&=& \frac{A^{F_N}_{Fib}(z)} { 1+ z^{F_{N-4}}-z^{F_{N-2}} }.
\eeq
The explicit form of the numerator $A^{F_N}_{Fib}(z)$ is 
given in appendix \ref{app:bunshi}.
The poles of the
$[\frac{F_N}{2}|\frac{F_N}{2}]$ Pade approximated function 
is given by zeros of the lacunary polynomial, 
\beq
1+ z^{F_{N-4}}-z^{F_{N-2}} =0.  
\eeq

\begin{figure}[htbp]
\begin{center}
\includegraphics[width=8cm]{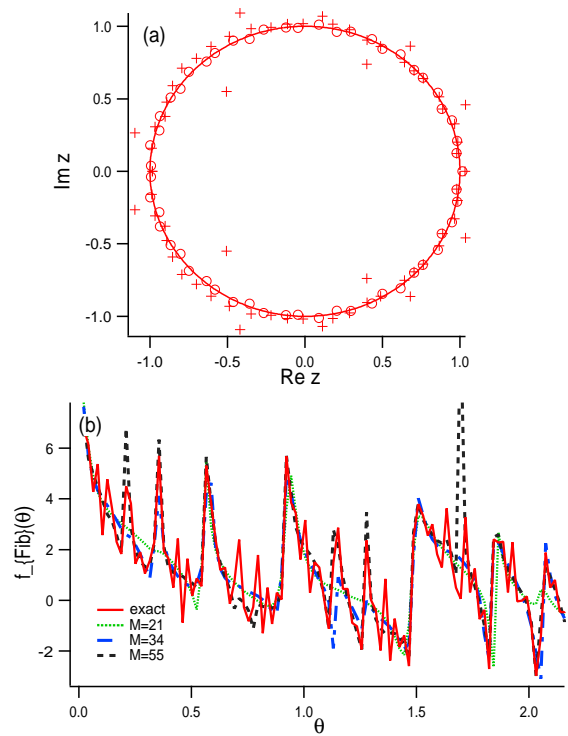}
\caption{
\label{fig:fig4} 
(Color online)
(a) Distribution of poles ($\bigcirc$) and zeros ($+$) 
of the $[55|55]$ Pade approximated function 
to the test function $f_{Fib}(z)$ with a natural boundary on $|z|=1$.
The unit circle is drawn to guide the eye.
(b)The Pade approximated functions 
$f_{Fib}^{[21|21]}(\theta)$, $f_{Fib}^{[34|34]}(\theta)$,
 $f_{Fib}^{[55|55]}(\theta)$ 
corresponding to $N=F_9=55$, $N=F_{10}=89$, $N=F_{11}=144$, respectively,
and the exact function $f_{Fib}(\theta)$ 
in the polar-form with $r=1.0$. (After Ref.\cite{yamada13}.)
} 
\end{center}
\end{figure}

In Fig.\ref{fig:fig4} the numerical result of 
the Pade approximation to $f_{Fib}(z)$ is shown.
The poles and zeros are plotted for 
the $[55|55]$ Pade approximation in Fig.\ref{fig:fig4}(a).
The poles and zeros  accumulate around $|z|=1$  
as increase of the order of the Pade approximation.
No pole appears inside the unit circle.
The original function is also well approximated 
by the $[56|56]$ Pade approximation.
(See Fig.\ref{fig:fig4}(b).)

\section{Natural boundary of random power series and the noise-effect on 
Pade approximation}
\label{sect:nb2}
In this section, we apply  
Pade Approximation to the random power series with a natural boundary with 
probability 1, 
and investigate how the approximation detect the singularity of the series.
In addition,   
we examine effect of noise on the coefficients of the power expansion for 
some test functions.
Some related theorems for the natural boundary of the function generated by the 
random power series are given in appendix \ref{app:noise1}.

\subsection{Random power series and natural boundary}
\begin{figure}[htbp]
\begin{center} 
\includegraphics[width=8.0cm]{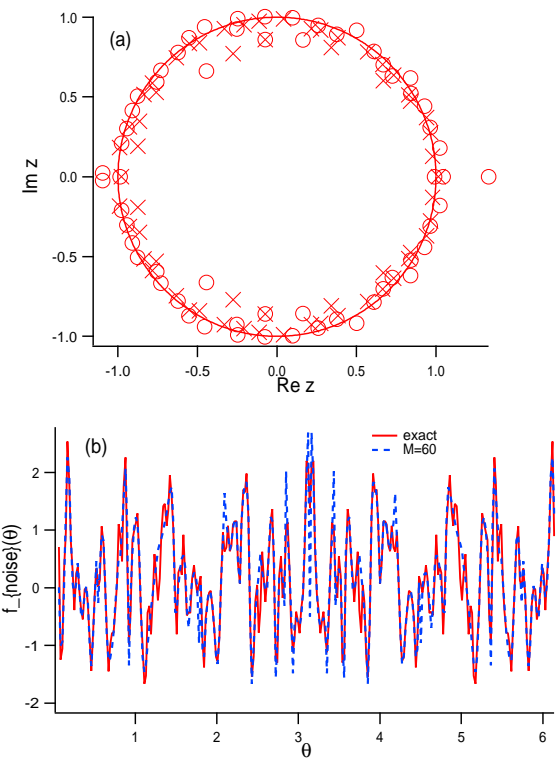}
\caption{
\label{fig:fig5}(Color online)
(a)Distribution of poles ($\bigcirc$) and zeros ($\times$) 
of the Pade approximated function $f_{noise1}^{[50|50]}(z)$ 
to a random power series $f_{noise1}(z)$ with $\epsilon=1$.
The unit circle is drawn to guide the eye.
(b)The Pade approximated function $f_{noise1}^{[50|50]}(\theta)$,
 and the exact function $f_{Fib}(\theta)$ 
in the polar-form with $r=1.0$.
}
\end{center} 
\end{figure}

Let us consider a random power series, 
\beq
 f_{noise1}(z) &=& \sum_{n=0}^{\infty} \epsilon r_n z^{n}. 
\label{eq:noisy-function}
\eeq
Here the sequence $r_n$  is $n-$independent random variable 
which take a value within $r_n \in [0,1]$, and 
 $\epsilon$ is the strength of the randomness. 
It is shown that, in general, the random power series  has a
natural boundary on the unit circle $|z|=1$ with probability one. 
Figure \ref{fig:fig5}(a) shows distribution of poles and zeros 
of the $[50|50]$ Pade approximated function
to $f_{noise1}(z)$.
Some pairs of poles and zeros are perfectly cancelled
 inside the circle $|z|=1$.
On the other hand, almost all the poles and zeros of the Pade approximated function
 assemble around the circle $|z|=1$ and not cancelled.
The pair of poles and zeros around the circle $|z|=1$
 is called "Froissart doublets", and it well correspond 
to the natural boundary
of $f_{noise1}(z)$.
The original function is also well approximated 
by the $[50|50]$ Pade approximation.
(See Fig.\ref{fig:fig5}(b).)

\begin{figure}[htbp]
\begin{center}
\includegraphics[width=8.0cm]{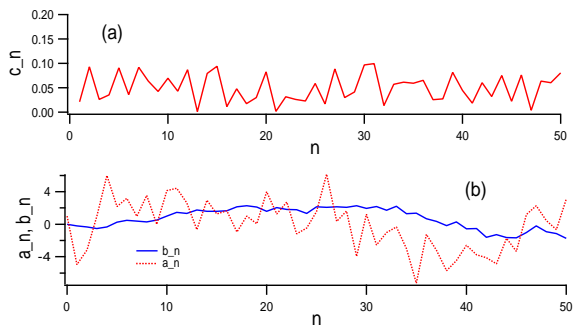}
\caption{(Color online)
\label{fig:abc}
(a)The coefficient $\{ c_n \}=\{ \epsilon r_n \}$ of
a truncated random power series $f_{noise1}^{[100]}(z)$
with $\epsilon=0.1$.
(b)The coefficients $\{ a_n \}$ and $\{ b_n \}$ of 
the Pade approximated function 
$f_{noise1}^{[50 \mid 50]}(z)$ to 
 $f_{noise1}^{[100]}(z)$.
} 
\end{center}
\end{figure}

 Figure \ref{fig:abc} shows an example of 
the coefficients $\{ c_n \}=\{ \epsilon r_n \}$  
of the random power series and 
 the coefficients $\{ a_n \}$ and $\{ b_n \}$  
of the $[50|50]$ Pade approximated function.
The fluctuation of 
the coefficient $\{ b_n \}$  that determines the poles 
of the Pade approximated function is smaller than those of $\{ a_n \}$
of the numerator.

The truncated random series is just random polynomial.
It is well-known that the distribution of the zeros 
of the random polynomials 
converges unit circle as increase of the order 
(Erdos-Turan type theorem) \cite{kac43,erdos50,peres05}. 
Accordingly, we can interpret that 
in the Pade approximated function to the random power series 
the distribution of poles and zeros also accumulate
around the unit circle when the order of the Pade approximation increases.

\subsection{Effect of noise on a function with a simple pole}
\label{subsect:noisy_pole}
In the following subsections, we investigate influences of noise
on the Pade approximation for some constructed noisy test functions 
as follows, 
\beq
f_{test+noise2}(z) &=& f_{test}(z) +f_{noise2}(z),
\eeq
where $f_{test}(z)=\sum_{n=0}^{\infty}a_n z^n$ and 
$f_{noise2}(z)=\sum_{n=0}^{\infty} \eps_n z^n$.
The $\{ \eps_n \}$ is a i.i.d. random variables
 within $[-\eps,\eps]$ where $\eps$ is the noise 
strength.
Essentially, $f_{noise2}(z)$ is the same as the random power 
series $f_{noise1}(z)$.
First of all, in this subsection, 
we consider a truncated function with a simple pole.
Note that if $a_n=C$(constant) and $\epsilon=0$, i.e., 
in noise-free case,
$f_{pole+noise}(z)=C\sum_{n=0}^{\infty}z^n =\frac{C}{1-z}$ 
with a simple pole at $z=1$. 
In Ref.\cite{baker75} by Baker, 
the noise effect is summarized as follows:
The $[M|M]$ Pade approximation has an unstable zero 
at the distance of order $\epsilon^{-1}$ from
the origin, 
and the other zeros make $(M-1)$ Froissart doublets (zero-pole pairs) 
with the zeros. 

Next, we consider a function 
\beq
f_{pole2+noise2}(z) &=& f_{pole2}(z)+ f_{noise2}(z), \\
&=& \sum_{n=0}^{\infty}(\frac{1}{2^n}+\epsilon_n)z^n, 
\eeq
with the noise strength $\eps <1 $.
Note that 
\beq
f_{pole2}(z)= \frac{2}{(2-z)},
\eeq
with a simple pole at $z=2$ to clearly show the shift of the poles 
of the approximated function due to the noisy series.

\begin{figure}[htbp]
\begin{center}
\includegraphics[width=7.0cm]{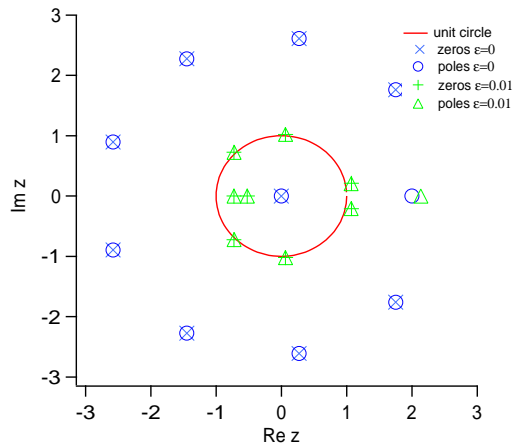}
\caption{
\label{fig:pole-noise}(Color online)
Distribution of poles ($\bigcirc$,$\bigtriangleup$) and zeros ($\times$,$+$) 
of the $[10|10]$ Pade approximated function 
$f_{pole2+noise2}^{[10|10]}(z)$
 with a stable pole at $z=2$
for noise strength $\epsilon=0$,$\epsilon=0.01$. 
The unit circle is drawn to guide the eye.}
\end{center}
\end{figure}

Figure \ref{fig:pole-noise} shows distribution of the poles and zeros of the 
$[10|10]$ Pade approximated functions.
It clearly shows the pole-shift by the noise effect.
In the noise-free case ($\eps=0$),
 a pole of the Pade approximation appears at $z=2$
and the other poles are cancelled with zeros 
(zero-pole ghost pairs). 
In a case when the relatively small noise ($\eps=0.01$)
is added, the poles and zeros move toward $|z|=1$ 
with making Froissart doublets, although a pole at $z=2$ is 
quite stable. It becomes impossible to detect the true pole at $z=2$
when the noise strength is relatively large ($\eps=0.1$), 
not shown in the Fig.\ref{fig:pole-noise}.

As a result, it is found that 
the locations of the ghost pairs are unstable for noise,  
and the residues for the poles are much smaller than
one corresponding to the true pole.
 We can guess that the proximity of the non-modal poles 
and zeros of the Pade approximated function can be understood
in a sense that the poles due to the noise 
need zeros to cancel with each other as
$\epsilon \to 0$.

\subsection{Effect of noise on a function with a branch cut}
\begin{figure}[htbp]
\begin{center}
\includegraphics[width=7.0cm]{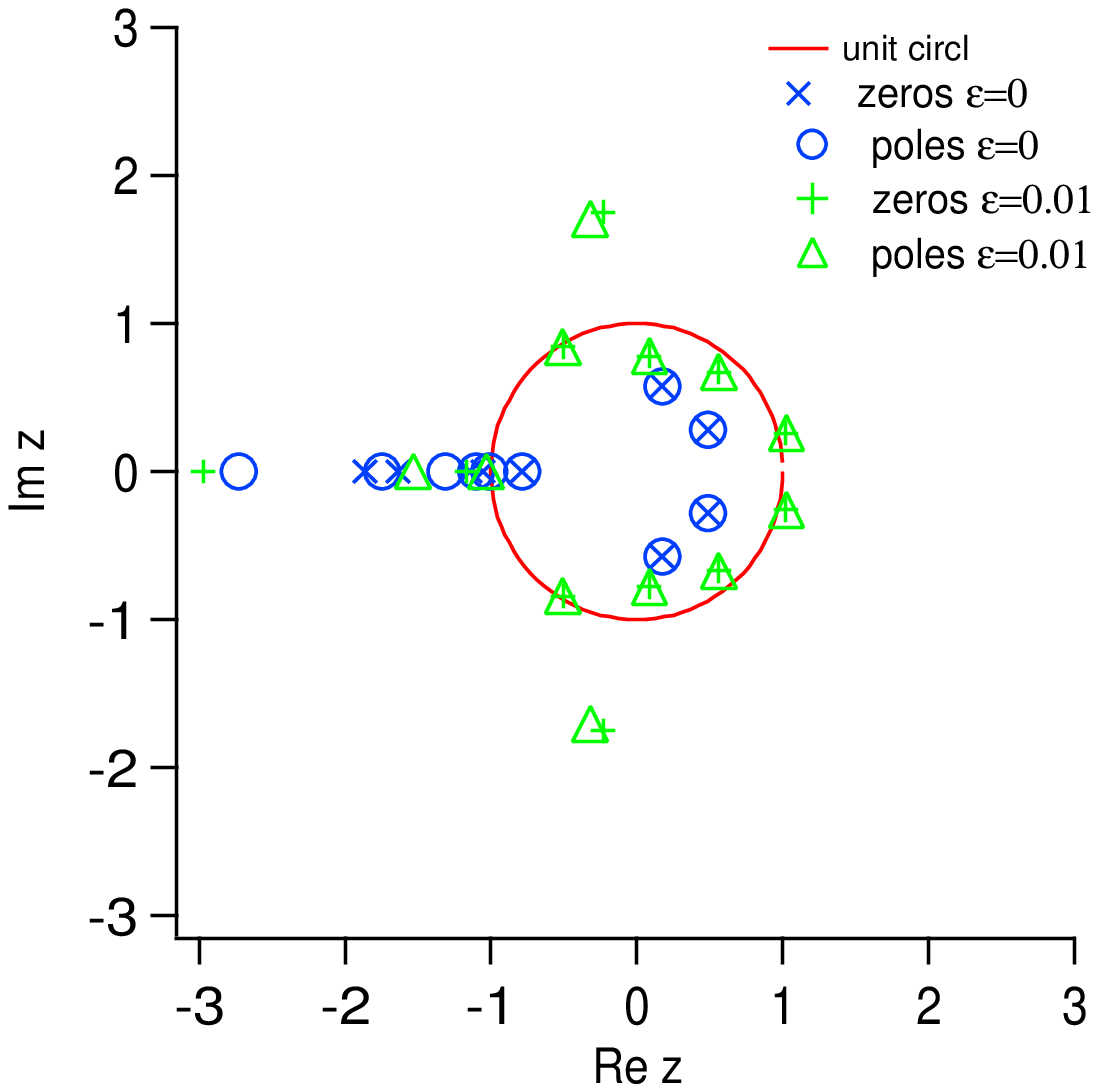}
\caption{
\label{fig:branch-noise}(Color online)
Distribution of poles ($\bigcirc$,$\bigtriangleup$) 
and zeros ($\times$,$+$) 
of the $[10|10]$ Pade approximated function 
$f_{branch1+noise2}^{[10|10]}(z)$
 with a brunch cut from $z=-\infty$ to $z=0$ 
for the noise strength $\epsilon=0$,$\epsilon=0.01$. 
The unit circle is drawn to guide the eye.} 
\end{center}
\end{figure}

We investigate the effect of the noise on functions with a branch cut.
First, let us consider a function 
\beq
f_{branch1}(z)=\sqrt{\frac{3+z}{1+z}},
\eeq
with an algebraic branch points at $z=-1$ and $z=-3$, and 
with the branch cut in $[-3,-1]$.
Distribution of the poles and zeros
of the Pade approximated function 
$f_{branch1+noise2}^{[10|10]}(z)$
is shown in Fig.\ref{fig:branch-noise}.
In a case with relatively small noise ($\eps=0.01$),
some poles make a line on the branch cut, and 
some poles and zeros move toward the unit circle $|z|=1$. 
It is impossible to detect the branch cut
when the noise strength is relatively large ($\eps=0.1$).

\begin{figure}[htbp]
\begin{center}
\includegraphics[width=8.0cm]{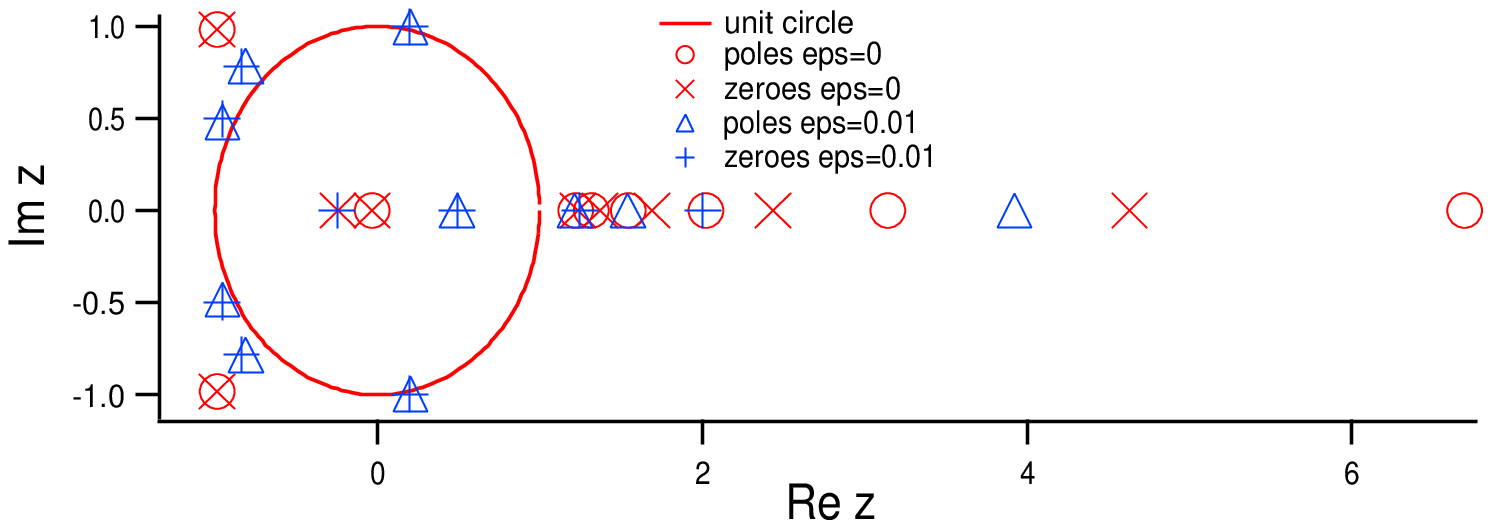}
\caption{
\label{fig:branch2-noise}(Color online)
Distribution of poles ($\bigcirc$,$\bigtriangleup$) and 
zeros ($\times$,$+$) 
of the $[10|10]$ Pade approximated function 
$f_{branch2+noise2}^{[10|10]}(z)$
 with a brunch cut from $z=6/5$ to $z=\infty$ 
for the noise strength $\epsilon=0$,$\epsilon=0.01$. 
The unit circle is drawn to guide the eye.} 
\end{center}
\end{figure}

Next, let us consider a function 
\beq
f_{branch2}(z)=\log (\frac{6}{5}-z),
\eeq
with a logarithmic branch point at $z=6/5$,
and with a brunch cut from $z=6/5$ to $z=\infty$.
The distribution of the poles and zeros
of the Pade approximated function $f_{branch2+noise2}^{[10|10]}(z)$
for the $f_{branch2}(z)$ with the noisy perturbation
is shown in Fig.\ref{fig:branch2-noise}. 
Some poles and zeros are making a line alternatively on the branch cut 
in the noise-free case ($\eps=0$).
It assemble around the unit circle $|z|=1$
with making Froissart doublets 
when the noise with strength $\eps=0.01$ is added.

\subsection{Effect of noise on a function with a natural boundary}
\begin{figure}[htbp]
\begin{center} 
\includegraphics[width=7.5cm]{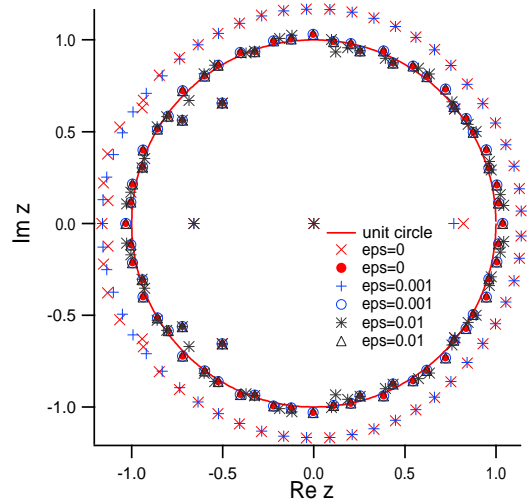}
\caption{ 
\label{fig:fig7}(Color online)
Distribution of poles ($\bullet,\bigcirc,\bigtriangleup$) and 
zeros ($\times+,*$) 
of the Pade approximated function $f_{Jac+noise}^{[50|50]}(z)$ 
for the lacunary series $f_{Jac+noise}^{[100]}(z)$ 
with a natural boundary on $|z|=1$. 
The noise strengths are $\epsilon=0$, $\epsilon=0.001$, 
$\epsilon=0.01$, respectively. 
The unit circle is drawn to guide the eye.
}
\end{center} 
\end{figure}

Figure \ref{fig:fig7} shows distribution of the poles and zeros 
of the $[50|50]$ Pade approximated
function to  
\beq
  f_{Jac+noise}(z) &=& f_{Jac}(z) + f_{noise2}(z),  
\eeq
which has a natural boundary on $|z|=1$.

In the noise-free case, 
 the pairs of poles and zeros of 
the Pade approximated function are perfectly cancelled
inside the unit circle $|z|=1$.
The other poles and zeros 
of the Pade approximated function
 assemble around the circle $|z|=1$ without cancellation.
In the relatively small noise case ($\eps=0.01$), 
the location of the poles is not significantly changed 
compared with the zeros shifted outside the unit circle 
due to the noise effect.
And, again, the  poles and zeros move toward $|z|=1$ 
with making Froissart doublets 
when the noise strength is relatively large ($\eps=0.1$).
It is closely related to a fact that 
fluctuation of the coefficients of the numerator 
of the Pade approximated function 
is much larger than those in the denominator, 
as seen in Pade approximation to  
the random power series in Fig.\ref{fig:abc}.
As a result, the singularity of the Pade approximated function 
to the function with a 
natural boundary is more sensitive to the noisy perturbation than 
those in the functions with the other type singularity such as 
simple poles and branch points.

It is very difficult to effectively distinguish 
whether the poles of the Pade approximation originated from  
the natural boundary on $|z|=1$ of the original function $f_{Jac}(z)$ or 
from the other natural boundary on $|z|=1$ generated 
by noisy series $f_{noise2}(z)$ or numerical errors.
Actually the round-off error effects on the distribution 
of the poles and zeros 
of the Pade approximated function.
Accordingly, to determine the expansion coefficients 
$c_n$ with adequate accuracy becomes 
very important in the numerical calculation.
This is a drawback of 
the Pade approximation when we use it for functions with unknown 
singularities.

\subsection{Numerical accuracy and spurious poles}
As we observed in the last subsection, 
the effect of rounding error and accuracy limit of computers
work in the numerical results of the Pade approximation.
As the result of accumulation of the round-off error, 
the "spurious poles" appear around the unit circle $|z|=1$
as the pole-zero pairs when the order of Pade approximation increases.
(We used a term "Froissart doublets" for  
the poles-zero pairs generated by random power series, conveniently, 
although we can not numerically distinguish it from the spurious poles
due to the round-off errors.
In the next section, we will discuss about the Froissart doublets again.)

However, we can roughly distinguish 
between true poles and the spurious poles 
by "residue analysis" of the Pade approximated function
because the spurious poles-zero pairs are unstable 
for the change of the order.
In this subsection, we try to investigate the residues 
of the Pade approximation for some
test functions.
 Up to now, the residue analysis has  been mainly used 
for performance comparison 
between the different algorithms of the Pade approximation 
of the same order \cite{gonnet11,gonnet12}.
On the other hand, it seems that the study 
by using the information of the residue analysis is 
still rare in the Pade approximation \cite{bessis96,gilewicz97}.

Generally, the rational polynomials of the diagonal Pade approximation can be 
uniquely identified 
by the poles $\{ z_k \}$ and 
the corresponding residues $A_k$ as follows:
\beq
\frac{Q_M(z)}{P_M(z)} = \sum_k^M \frac{A_k}{z-z_k},
\eeq
where the residues are given by 
\beq
A_k = \frac{Q_M(z_k)}{\prod^M_{j(\not = k)}(z_k-z_j) }.
\eeq
\noindent　
Here, we investigate the convergence property of 
the magnitude of residues $|A_k|$ 
 arranged in descending order.

\begin{figure}[htbp]
\begin{center}
\includegraphics[width=6cm]{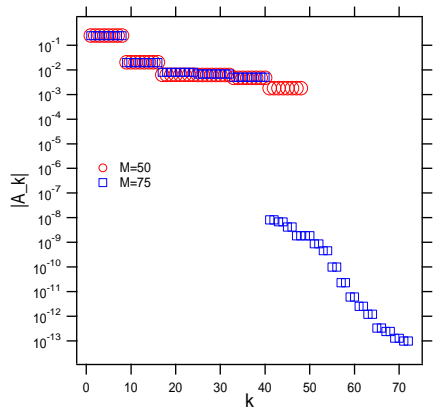}
\caption{
\label{fig:resi1}(Color online)
Absolute values of the residues $|A_k|$ of the $[50|50]$ and
 $[75|75]$ Pade approximated
functions to the test function $f_5(z)$ without noise. 
The $|A_k|$ are arranged in descending order. 
}
\end{center}
\end{figure}

Figure \ref{fig:resi1} shows  
the absolute value of the residues $|A_k|$ of some Pade approximated
functions to the test function $f_5(z)$, which are arranged in descending order.
(Note that it is noise-free cases.)
The distribution of the poles and zeros of the Pade approximated
functions is given in Fig.\ref{fig:tan}.
In a case of $M=50$, the magnitude of the all residues $|A_k|$ 
is larger than $O(10^{-3})$,
which correspond to the relevant poles arranged radially in eight directions
from the true poles.
On the other hand, in a case of $M=75$, the spurious poles appear and distribute
around the unit circle $|z|=1$. (See Fig.\ref{fig:tan}(b).)
It is found that the absolute values of the residues  
corresponding the spurious poles are several order of the magnitude smaller 
than the relevant poles.

\begin{figure}[htbp]
\begin{center}
\includegraphics[width=8cm]{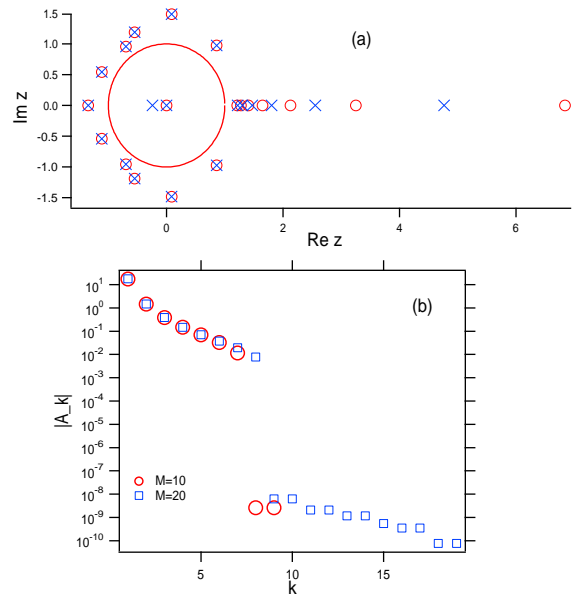}
\caption{(Color online)
\label{fig:resi2}
(a)Distribution of poles ($\bigcirc$) and zeros ($\times$) 
of the $[20|20]$ Pade approximated function $f_{branch2}^{[20|20]}(z)$.
The unit circle is drawn to guide the eye. 
(b)Absolute values of the residues $|A_k|$ of the $[10|10]$ and 
$[20|20]$ Pade approximated
functions to the test function $f_{branch2}(z)$ without noise. 
The residues are arranged in descending order. 
}
\end{center}
\end{figure}

Distribution of poles and zeros 
of the Pade approximated function $f_{branch2}^{[20|20]}(z)$
to the test function $f_{branch2}(z)$ is shown in Fig.\ref{fig:resi2}.
The stable poles and zeroes are lined on $[6/5,\infty]$, 
and the spurious poles appear around $|z|=1$.
The magnitude of the residues of the spurious poles is also enormously
small compared with those of the stable poles remained 
as increase of the order of the 
Pade approximation.

\begin{figure}[htbp]
\begin{center}
\includegraphics[width=7cm]{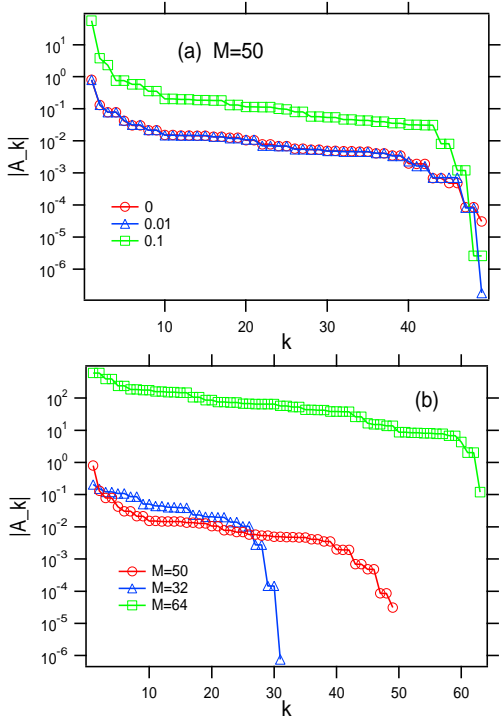}
\caption{(Color online)
\label{fig:resi3}
(a)Absolute values of the residues $|A_k|$ of the $[50|50]$ Pade approximated
functions to the noise added test function $f_{Jac+noise}(z)$ with 
the strength $\eps=0,0.01,0.1$. 
(b)Absolute values of the residues $|A_k|$ of the Pade approximated
functions $f^{[50|50]}_{Jac}(z)$, $f^{[32|32]}_{Jac}(z)$, 
$f^{[64|64]}_{Jac}(z)$ 
to the test function $f_{Jac}(z)$ without the noise. 
The residues are arranged in descending order. 
}
\end{center}
\end{figure}

Figure \ref{fig:resi3} is also the result of the residues analysis 
for the Pade approximated function to the test function $f_{Jac}(z)$
with a natural boundary on the unit circle $|z|=1$.
In the $[50|50]$ Pade approximated function, 
 the magnitude of the residues $|A_k|$ is shown in changing 
the noise strengths $\eps=0,0.01,0.1$
corresponding to poles-zeros distribution in Fig.\ref{fig:fig7}.

In the small noise case ($\eps=0.01$), 
the results of the residue analysis for $f_{Jac+noise2}(z)$ 
is almost same as the noise-free case ($\eps=0$), and 
in the case with relatively strong noise ($\eps=0.1$),
the noise shift the magnitude of the residues larger value. 
In addition, the result of the residue analysis of the noise-free cases 
for some different order of the Pade approximation
is shown in Fig.\ref{fig:resi3}(b).
We should have in mind that the order is important   
when we apply Pade approximation to the lacunary power series
because we should not take the order of the approximation 
in the gap of the series.

\section{Froissart doublets}
\label{sect:nb-random2}
The problem of constructing the $Z-$transform $Z(z)$ of a finite time-series 
is a standard problem in Mathematics 
\cite{bessis96,stahl97,gilewicz97,bessis09,antonio09}.
 For example, it is shown that for a sum of oscillating damped signals, 
the $Z-$transform associated with the
 time-series can be characterized by a sum of the poles 
of the Pade approximated function. 
The position of each pole is simply linked to the damping factor
and the frequency of each of the oscillators. 
Also, it is important to note that all these poles lie strictly outside the
unit circle because it corresponds to the damping
\cite{bessis96,stahl97,gilewicz97,bessis09}.
In addition, we will consider quasianalyticity property 
of the random power series by the residue analysis of 
the Pade approximation.

\subsection{Noise attractor}
In signal processing, 
we can use fact that the poles and zeros of the Pade approximated function 
to the noisy series distribute around the unit circle $|z|=1$ 
when we remove the noise from the observed data 
through $Z-$transform and/or Fourier transform of the data.
Let a sequence $\{s_0,s_1,....,s_n,...\}$ as a sample signal 
without noise.
Then we define the $Z-$transform of the sequence as,
\beq
Z(z)=\sum_{n=0}^N s_n z^{n}. 
\eeq
The function $Z(z)$ is analytic interior of $|z|<1$ 
if the number of signal $N$ is finite \cite{def}.
Note that Discrete Fourier transform is a special case of 
the $Z-$transform.

Next, in the concrete, let us consider a signal sequence 
in $t \in [0,T]$ consisting of the superimposed damping oscillators 
as, 
\beq
s_k = \sum_{\ell} A_{\ell} e^{i\omega_{\ell} \frac{k}{N} T}, k=0,1,...,N-1,
\eeq
where $A_{\ell}$ is the amplitude of the $\ell-$th oscillator, and 
$\omega_{\ell} =2\pi f_{\ell} +i \alpha_{\ell}$.
 Here, $f_{\ell}$ and $\alpha_{\ell}$ are the frequency and the damping factor 
of the $\ell-$th oscillator. 
Then, the $Z-$transform is 
\beq
Z(z)&=& \sum_{n=0}^\infty s_n z^{n} \\
&=& \sum_{n=0}^\infty \sum_{\ell} A_{\ell} e^{i\omega_{\ell} \frac{n}{N}
T } z^{n}  \\
&=& \sum_{\ell} \frac{A_{\ell}}{1-zz_{\ell}},
\eeq
where we take a limit $n\to \infty$ keeping $T/N$, and 
$z_{\ell} \equiv e^{i\omega_{\ell} \frac{T}{N}}$.
Accordingly, the singularity of $Z(z)$ appears 
as the poles at $z=z_{\ell}^{-1} \equiv e^{-i\omega_{\ell}
\frac{T}{N} } $ outside the unit circle $|z| > 1$ and the residue is 
$Res(z_{\ell}^{-1})=z_{\ell}^{-1}A_{\ell}$.

On the other hand, let us consider a noise-added 
sequence $\{S_0,S_1,....,S_n,...\}$. 
Then, the Froissart pointed out that there are two-types of the poles;
stable poles and unstable poles 
when we apply the diagonal Pade approximation to the unknown data set. 
In general, the $Z-$transform $Z(z)=\sum_{n=0}^N S_n z^{n}$
of the noisy sequence has a natural 
boundary on the unit circle $|z|=1$ with probability 1. 
In fact, the poles and zeros (Froissart doublets) 
of the Pade approximated function often distribute 
around the unit circle 
when the numerical error and/or noise are mixed 
into the Taylor series of the analytic functions,
as seen in the last section.
That is to say,  we sometimes call 
the unit circle $|z|=1$ {\bf noise attractor}  
in a sense that the poles and zeros are attracted to the circle 
as the Froissart doublets.
Accordingly, it is found that 
Pade approximated function for the function $Z(z)$ 
has stable poles associated with the 
damping modes and unstable spurious poles associated 
with the noisy fluctuation.
After elimination of the spurious poles around the noise attractor 
from the noisy-sequence 
we can reconstruct the noise-free sequence consisting of the stable poles 
located in the domain $|z|>1$.  
Another remarkable feature of the non-modal poles
is that the absolute value of the Cauchy residues 
associated with them are usually 
much smaller than those associated with true poles. 

\subsection{Random power series and quasianalytic function}
Weierstrass defined the analytic function by 
direct analytic continuation of function.
Then, apparently the analytic continuation is impossible 
beyond the natural boundary 
even if we can uniquely define the function and it is analytic 
outside the analytic domain. 
Borel and Gammel extended the narrow condition for the analyticity 
and gave a definition of 
{\it quasianalytic functions} \cite{Luzin98a,Luzin98b}.  
Gammel conjectured the following for 
the random power series \cite{gammel73,bessis96}.

{\bf Gammel conjecture(1973):}
The random power series belongs to
the Borel class of quasianalytic functions as the following form,
\beq
  f_{Gammel}(z)=\sum_{k=0}^{\infty} \frac{B_k}{1-w_kz},
\label{eq:gammel}
\eeq
where $w_k=e^{i 2\pi X_k}$ and $\{X_k\}$ are real numbers 
in the interval in $X_k \in [0,1]$, and $B_k$ decrease rapidly with $k$.
Then natural boundary in the Weierstrass's sense can be crossed.
■

The function Eq.(\ref{eq:gammel}) is a simple example 
that poles are densely-distributed on the unit circle.  
Then the convergence property of the sequence $|B_k|$ is 
important for the analyticity of the function.  
Carleman proved that $f_{Gammel}(z)$ is quasi-analytic 
if $B_k$ satisfies the following condition 
\beq
|B_k| < Ce^{-k^{1+c}}, c > 0.
\label{eq:gammel-condition}
\eeq
This is Carleman-class of quasianalytic functions.
See Gammel's paper \cite{gammel73} for the details.
Moreover, Gammel and Nuttall proved that the quasianalytic functions 
can be exactly approximated by the Pade approximation \cite{gammel73}.

{\bf Gammel-Nuttal Theorem(1973):}
If $B_k$ in Eq.(\ref{eq:gammel}) satisfies the condition 
(\ref{eq:gammel-condition}), and $|\omega_k|=1$,
then, the sequence of $[N+J|N]$ Pade approximation to the $f_{Gammel}(z)$ 
converges in measure to the function $f_{Gammel}(z)$ as $N \to \infty$
in any closed, bounded region of the complex plane,
where $J$ is a natural number that equals to $N$ or less. 
■

\begin{figure}[htbp]
\begin{center}
\includegraphics[width=6.5cm]{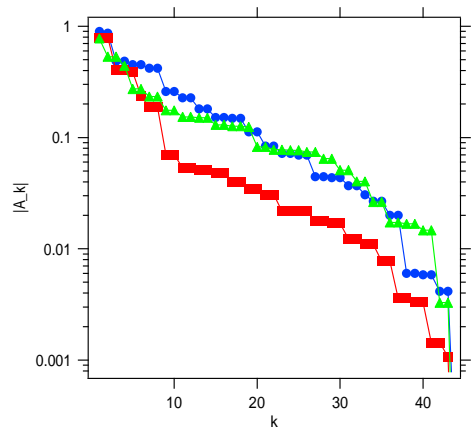}
\caption{
\label{fig:resid}(Color online)
Absolute values of the residues $|A_k|$ of the Pade approximated
functions $f^{[45|45]}_{noise2}(z)$ 
to three samples of the truncated random 
power series of order $N=90$, and $\eps=1$.
The $|A_k|$ are arranged in descending order. 
}
\end{center}
\end{figure}

{\it Is the Gammel conjecture true?}
We try to examine the validity of the Gammel conjecture 
by applying residue analysis of the Pade approximated function 
to the random power seres $f_{noise2}(z)$.
Figure \ref{fig:resid} shows the absolute values of 
the residues $|A_k|$ of 
the Pade approximated
functions $f^{[45|45]}_{noise}(z)$ to 
three different samples in descending order.
$|A_k|$  roughly 
exponentially decreases with respect to $k$ as, 
\beq
 |A_k| \sim exp(-\beta k), 
\eeq
where $\beta$ is the decay exponent.
It shows exponential  decay (or faster), and, at the first face,  
supports the Gammal conjecture.

\begin{figure}[htbp]
\begin{center}
\includegraphics[width=7cm]{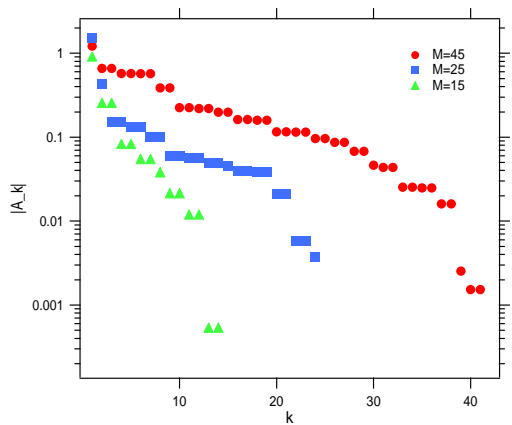}
\caption{
\label{fig:resid2}(Color online)
Absolute values of the residues $|A_k|$ of the Pade approximated
functions $f^{[45|45]}_{noise2}(z)$, $f^{[25|25]}_{noise2}(z)$, 
$f^{[15|15]}_{noise2}(z)$
 to a sample truncated random 
series, and $\eps=1$.
The $|A_k|$ are arranged in descending order. 
}
\end{center}
\end{figure}

However, it is not nearly so simple.
We should check the stability of the exponential-like decay of the 
magnitude of the residues by changing the order of the Pade approximation.
Figure \ref{fig:resid2} shows the result for the three different orders 
$M=15$, $M=45$, $M=55$.
It express an indication that the decay exponent $\beta$ does not 
converge a positive certain value.
It seems that the exponent behaves $\beta \to 0$ as a limit $M \to \infty$. 
On the other hand, 
if we directly apply the Pade approximation to the quasianalytic function 
$f_{Gammel}(z)$ with $B_k=\e^{-k}$, the exponent $\beta$ is stable for 
changing the order of the Pade approximation. (See appendix \ref{app:residue}.)
These facts suggest that th random power series is 
not belong to Carleman-class of quasianalytic function
although it has a natural boundary on the unit circle, 
and it has the form (\ref{eq:gammel}).
As a result, we can say that no optimism is warranted on 
the Gammel conjecture.

How does the residue analyses of the Pade approximation 
to analyticity and/or 
quasianalyticity of unknown function work?
It is an interesting and future problem.

\section{Summary and Discussion} 
In the present paper 
we numerically examined the effectiveness of the Pade approximation 
to some test functions with branch point, essential singularity, 
and natural boundary
by watching the singularities of the Pade approximated functions. 
For the functions with a branch cut, the poles and zeros of 
the Pade approximated function are lined 
along the true branch cut. 
The poles and zeros are distributed around the true natural boundary 
if the original test function has a natural boundary. 
In addition, we gave explicit the Pade approximated functions 
to some lacunary power series which are useful to check the numerical result. 
It was shown that, in particular, 
the distribution of poles and zeros 
of the Pade approximated function 
to lacunary power series and the random power series 
accumulated around the unit circle when the order 
of the approximation increases.

We often suffer from the difficulty 
to distinguish whether or not the poles of the Pade approximation 
are intrinsically 
originated from the natural boundary of the original power 
series, because the numerical errors contained in the expansion 
coefficients also yields a false natural boundary.
Therefore, the expansion coefficients with 
adequate numerical accuracy are necessary when we apply 
the Pade approximation to functions with unknown singularities.
 
Furthermore, the residue calculus of the Pade approximated function 
is useful when we detect the singularity of the original power series 
from the asymptotic behavior of the truncated series. 
It is useful also for estimating the accuracy of the approximation. 
As a result, the residue calculus suggests that 
the random power series does not obey Gammel conjecture, 
that is, it does not belong to Borel-class of quasianalytic functions.

We finally remark that the most serious 
problem to be improved is the numerical accuracy 
due to the limitation of the order in the 
Pade approximation when we use it 
for detecting unknown singularities of wave functions
in quantum physics\cite{yamada13}.

\appendix

\section{General recursion relation}
\label{app:recursion}
We can construct a power series that has some pole-type singularities 
in the following form
\beq
\frac{dz^2+ez+f}{az^2+bz+c} = \sum_{n=0}^{\infty} a_n z^n,
\label{eq:pole}
\eeq
where $a,b,c,d,e,f$ are real and $c \neq 0$ for simplicity.
Then the coefficients $\{ a_n \}$ can be obtained by 
rearranging and comparing with the coefficients of the both sides
 in the same order as follows,
\beq
dz^2+ez+f = (ba_0z+ca_0+ca_1z) \\ \nn
+ \sum_{n=2}^{\infty}
(aa_{n-2}+ba_{n-1}+ca_{n} )z^n.
\eeq

As a result, the power series with the pole-type singularities 
can be constructed 
by the recursion relation 
\beq
a_k=-\frac{b}{c}a_{k-1}-\frac{a}{c}a_{k-2}, k\geq 2,
\eeq
with $ca_0=f$, $ba_0+ca_1=e$, $aa_0+ba_1+ca_2=d$.

It becomes Fibonacci sequence 
when we set $a_0=0, a_1=1$, $a_k=a_{k-1}+a_{k-2}$.

\section{Random polynomial}
\label{app:random-polynomial} 
The following theorems concerning the random power series are well-known.

{\bf Erdos-Turan Type theorem(1950):}
Let us define a polynomial
\beq
f(z)=\sum_{n=0}^N a_n z^{n},
\eeq
where coefficients $a_n$ are randomly distributed 
and $a_0 a_N \neq 0$, for simplicity.
Then the zeros of the random polynomial cluster uniformly around the 
unit circle $|z|=1$ if "size of the truncated series" $L_N(f)$ is 
small compared to the order $N$ of the polynomial, 
where 
\beq
L_N(f) = \log \left( \frac{\sum_{n=0}^{N} |a_n|}{\sqrt{|a_0a_N|} } \right).
\eeq 
■

Note that this theorem also hold for the polynomials with deterministic 
coefficients $a_n$ such as Newman type polynomial having coefficients in the 
sets $\{ 0,1 \}$ or $\{ 0, \pm1 \}$. 

{\bf Peres-Virag's theorem(2005):}
Let us $\{ a_n \}$ are i.i.d. Gaussian-type random variables, 
then the distribution $K(z)$ of the complex zeros $\{ z_k \}$ of the power series
\beq
f(z)=\sum_{n=0}^\infty a_n z^{n}, 
\eeq
is 
\beq
K(z_k)=  \frac{1}{\pi} \frac{1}{(1-|z_k|^2)^2}. 
\eeq
■

\section{Some Gap Theorems of Lacunary Power Series}
\label{app:lacunary-thorems}
Weierstrass considered the analyticity of the power series,
\beq
 f(z)=\sum_{\nu=0}^\infty a_\nu z^{b^\nu},  b \in N, b \neq 1,
\label{eq:wei}
\eeq
where $a_\nu$ is a positive number. 
In the main text, we set $a_\nu=1$, $b=2$ for $f_{Wei}(z)$.
Then, it is proved that
 the function (\ref{eq:wei}) has a natural boundary on the unit circle $|z|=1$
if the convergence radius of the function is unity based on  
the following theorems for the lacunary power series.

{\bf Hadamard-Barck's gap theorem(1892):} 
Let us be 
\beq
 f(z)=\sum_{\nu=0}^\infty a_\nu z^{\lambda_\nu}, 
\eeq
where $a_\nu$ is a positive number, and $\{ \lambda_\nu \}$ denote
a strictly increasing sequence of the natural numbers, 
satisfying a inequality $q\lambda_\nu \leq \lambda_{\nu+1}$ for $q>1$.
Then the function $f(z)$ has 
a natural boundary on the unit circle $|z|=1$. ■

{\bf Fabry's gap theorem(1899):}
Power series 
\beq
f(z)=\sum_{\nu=0}^\infty a_\nu z^{\lambda_\nu},
\eeq 
with radius of convergence $R=1$ 
has a natural boundary on the unit circle $|z|=1$, provided it is 
Fabry series, i.e.  
\beq
 \lim_{\nu \to \infty} \frac{\lambda_\nu }{\nu} = \infty. 
\eeq
■

\section{Numerators of diagonal Pade approximations 
to $f_{Jac}(z)$ and $f_{Fib}(z)$}
\label{app:bunshi}
The diagonal Pade approximation to the truncated 
lacunary power functions $f_{Jac}(z)$ and $f_{Fib}(z)$
 can be exactly executed as given in the main text.
The numerators $A_{Jac}^N(z)$ and $A_{Fib}^{F_N}(z)$  of the 
Pade approximated functions can be given as follows,
\beq
A^N_{Jac}(z) &=& z+2z^2  \\ \nn
&+& 2\sum_{n=2}^{N-1}z^{H_n} (z+z^2+\sum_{k=1}^{n-2}
z^{H_{k+2}}),
\eeq
where $H_n=2^{n-1}$.

Numerator of the diagonal Pade approximated function 
to $f_{Fib}(z)$ is 
\beq
A^{F_N}_{Fib}(z) &=& S_{N-4}(z)  \nn \\
&+& [S_{N-8}(z)+z](f_{N-4}(z)-f_{N-2}(z))  \\  
&+& [2f_{N-3}(z)+2f_{N-2}(z)+f_{N-3}(z)f_{N-6}(z)], \nn
\eeq
where $S_{L}(z)=\sum_{k=0}^{L} f_{k}(z)$, $f_k(z)=z^{F_k}$. 
$F_N$ means $N$th Fibonacci number, and we set 
$F_{-1}=F_{-2}=....=0$.

We have inductively obtained above results by means of 
MATHEMATICA.

\section{Residue analysis for Carleman class of quasianalytic functions}
\label{app:residue}
\begin{figure}[htbp]
\begin{center}
\includegraphics[width=6.5cm]{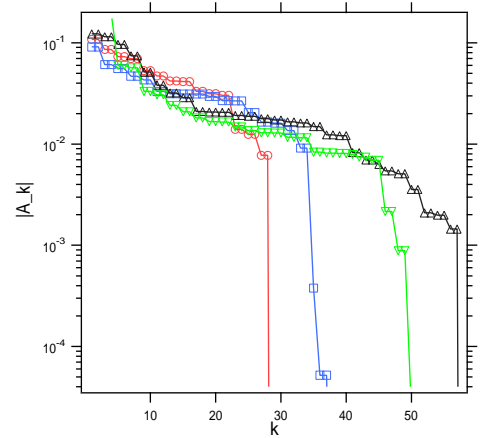}
\caption{
\label{fig:quasi-analy1}(Color online)
Absolute values of the residues $|A_k|$ of the Pade approximated
functions to a truncated Carleman function $f_{Carleman}(z)$ 
 of order $M=15,25,45$, which 
is artificially constructed by Eq.(\ref{eq:quasi-example}).
They are arranged in descending order in each case. 
}
\end{center}
\end{figure}

In this appendix, we give a direct result of residue analysis 
for "Carleman-class" 
of the quasianalytic functions, for comparison with 
the other residue analysis 
in the main text.
We apply 
the Pade approximation to the   
qusiperiodic function $f_{Carleman}(z)$ of the Carleman-class, 
which is artificially constructed by a set of the poles $\{ z_k \}$ as follows;
\beq
 f_{Carleman}(z) &=& \sum_{k=1}^{K} 
 ( \frac{1}{1-z_kz} +  \frac{1}{1-z_k^{*}z} )
e^{-k}                  \\
&=& 2\sum_{n=0}^{\infty} \sum_{k=1}^{K} 
e^{-k} \cos(2\pi X_k n) z^n,
\label{eq:quasi-example}
\eeq
where we set the poles 
at $z_k=\exp(\pm 2\pi i X_k)$ ($k=1,2,...,K$) on the unit circle. 
$\{ X_k \}$ are i.i.d random variables 
in the interval $X_k \in [0.1]$ and we take $K=100$. 
Figure \ref{fig:quasi-analy1} shows  
the absolute values of the residues $|A_k|$ of the Pade approximated
functions of order $M=15$, $M=25$, $M=45$ to $f_{Carleman}(z)$.
They are arranged in descending order. 

As a result, it seems that $|A_k|$ exponentially decreases 
with a stable exponent, regardless of the order of the Pade approximation.
This supports that 
certainly, the Pade approximation is applicable to the quasianalytic 
functions in the Gammel conjecture as given in Gammel-Nuttal's theorem.
The Pade approximation for the quasianalytic function converges to 
the function even outside the unit circle.
It should be also noted that in all cases the tails of $|A_k|$ are 
rapidly decay because 
the "truncated" series are essentially analytic functions.

\section{Some Results for Natural Boundary in Noisy Series}
\label{app:noise1}
In this appendix some theorems for the random 
power series are given. 
See, for example, Ref.\cite{remmert10} for the proofs.

{\bf Steinhaus's theorem(1929):}
Suppose that the power series,
\beq
f(z)=\sum_{n=0}^{\infty} a_n z^n,
\label{eq:ap:power-series}
\eeq
has radius of convergence $R=1$.
Let us $X_0, X_1,..., X_n$ be a sequence of i.i.d. random variables 
 in the interval $X_i \in [0,1]$.
Then, with probability one, the random power series
\beq
f_{Steinhaus}(z) &=& \sum_{n=0}^{\infty} a_n w_n z^n,
\eeq
has a natural boundary on $|z|=1$, where $w_k=e^{i 2\pi X_k}$. 
■

{\bf Paley-Zygmund's theorem(1932):}
Suppose that the power series (\ref{eq:ap:power-series}) 
has the radius of convergence 1.
Let us $r_0, r_1,..., r_n,...$
be a sequence of binary stochastic variables 
taking $-1$ or $1$ with equal probability.
Then, with probability one, the random power series
\beq
f_{P-Z}(z) &=& \sum_{n=0}^{\infty} r_n z^n,
\eeq
has a natural boundary on the unit circle $|z|=1$.
■

The similar theorems can hold for 
random power series $\sum_{n=0}^{\infty} r_n z^n$
with a sequence of stochastic variables
obeying i.i.d. in the interval $r_i \in [-1,1]$ or $r_i \in [0,1]$
\cite{kahane85}. The more generalized version has been given in 
the following form \cite{breuer11}.

{\bf Breuer-Simon's theorem(2011):}
Suppose that the power series (\ref{eq:ap:power-series}) 
has the convergence radius 1.
 Then for a.e.
$\omega$, $f(z)=\sum_{n=0}^{\infty}a_n(\omega) z^n$
has a strong natural boundary on $|z|=1$
if the $a_n(\omega)$ be a stationary,
ergodic, bounded, nondeterministic process.
■

\section*{Acknowledgments}
This paper was partially written 
for  "International Symposium of Complexified Dynamics, 
Tunnelling and Chaos" held on 2005 in Kusatsu.
This work is partly supported by Japanese people's tax via MEXT,
and the authors would like to acknowledge them.
They are also very grateful to Dr. T.Tsuji and 
to all concerned in Koike memorial
house for using the facilities during this study.



\begin{thebibliography}{00}

\bibitem{baker70} G.A. Baker and J.L. Gammel,
{\it The Pad\'e Approximation in Theoretical Physics},
(Academic Press: New York, 1970).

\bibitem{baker75} George A. Baker Jr,
{\it Essentials of Pad\'e Approximants} (Academic Press, 1975).


\bibitem{baker96} George A. Jr Baker and Peter Graves-Morris,
{\it Pad\'e Approximants 2nd Edition}, (Cambridge University Press, 1996).





\bibitem{sasagawa81}
F. Sasagawa, {\it Scattering Theory}(Syoukabou, 1991）
(in Japanese).

\bibitem{stanly71}
H. Stanly, {\it Introduction to Phase Transitions and Critical Phenomena}
 (Clarendon Press, Oxford, 1971). 

\bibitem{yang52}  
C.N. Yang and T.D. Lee, 
Phys. Rev. {\bf 87} (1952) 404.

\bibitem{nickel99}  
B. Nickel, 
 J. Phys.A {\bf 32}  3889(1999): 
 J. Phys. A {\bf 33} 1693(2000) .

\bibitem{kubo85} 
R. Kubo, M. Toda, N. Hashizume, and N. Saito, 
{\it Statistical Physics I, II} (Springer-Verlag, Berlin, 1983, 1985).

\bibitem{macoy01}  
B. M. McCoy, 
arXiv:cond-mat/0103556v1

\bibitem{bessis96}  
D. Bessis, 
J. Comput. Appl. Math. {\bf 66}  85-88(1996).

\bibitem{stahl97}  
H. Stahl,
J. Comput. Appl. Math. {\bf 86}  287-296(1997).

\bibitem{gilewicz97}  
J. Gilewicz and M. Pindor, 
J. Comput. Appl. Math. {\bf 87}(1997). 


\bibitem{bessis09}  
D. Bessis and L. Perotti,
J. Phys. A {\bf 42}, 365202(2009) .

\bibitem{antonio09}  
L.A. Barbosa Coelho and L.A. Baccala,
Pade approximations as a Modal Identification Technique,
in Proceedings of the IMAC-XXVII held 
February 9-12, 2009 Orlando, Florida USA.

\bibitem{berrretti90} A. Berretti and L. Chierchia, 
Nonlinearity {\bf 3}, 39(1990). 




\bibitem{falcolini92} C. Falcolini and R. Llave,  
J. Stat. Phys. {\bf 67}, 645(1992).  



\bibitem{llave94b} R. Llave and S. Tompaidis, 
Physica D {\bf 71}, 55(1994). 



\bibitem{berretti95} A. Berretti and S. Marmi, 
Chaos, Solitons and Fractals, {\bf 5}, 257(1995).　

\bibitem{berretti01} A. Berretti and C. Falcolini, and G. Gentile, 
Phs. Rev. E {\bf 64}, 015202-1(2001).　


\bibitem{korner93} 
T.W. Korner, {\it Exercises for Fourier Analysis}
(Cambridge University Press 1993).


\bibitem{remmert10} 
R. Remmert, {\it Classical Topics in Complex Function Theory},
(Springer New York; 1st Edition 2010). 

\bibitem{breuer11}  
J. Breuer and B. Simon,
Advances in Mathematics
{\bf 226}, 4902-4920(2011).
arXiv:1002.0823v2 [math.CV].

\bibitem{knill10}
O. Knill and J. Lesieutre,
"Analytic continuation of Dirichlet series with almost 
periodic coefficients",  
Complex Analysis and Operator Theory 
{\bf 6}, 237-255(2010). 
arXiv:0811.1362.

\bibitem{costin}  
O. Costin and M. Huang,
Behavior of lacunary series at the natural boundary,
Advances in Mathematics {\bf 222}, 1370-1404(2009).



\bibitem{press88}
W. H. Press, S. A. Teukolsky, W.T. Vetterling and B.P. Flannery,
{\it Numerical Recipes in C} (Cambridge University Press, 1988);
W. H. Press, S. A. Teukolsky, Computers in Physics, {\bf 6}, 82(1992).


\bibitem{gonnet11}
P. Gonnet, R. Pachon, and L. N. Trefethen,
Electronic Transactions on Numerical Analysis
{\bf 38}, 146-167(2011).

\bibitem{gonnet12}
P. Gonnet, S. Guttel, and L. N. Trefethen,
{\it Robust Pade Approximation via SVD},
SIAM Review {\bf 55}, 101-117(2013).









\bibitem{kac43}
M. Kac,
Bull. Amer. Math. Soc. {\bf 49}, 314-320(1943).

\bibitem{erdos50}
P. Erdos and P. Turan, 
Annals of Mathematics {\bf 51}, 105-119(1950).

\bibitem{amoroso96}
F. Amoroso and M. Mignotte, 
Ann. Inst. Fourier {\bf 46}, 1275-1291(1996).

\bibitem{odlyzko93}
A. Odlyzko and B. Poonen, 
Enseign. Math. {\bf 39}, 317-348 (1993).

\bibitem{simon04a}
B.Simon, 
{\it Orthogonal Polynomials on the Unit Circle}, 
part 1:Classical Theory, 
(American Mathematical Society  2004). 

\bibitem{simon04b}
B.Simon, 
{\it Orthogonal Polynomials on the Unit Circle}, 
part 2:Spectral Theory, 
(American Mathematical Society  2004). 

\bibitem{peres05}
Y. Peres and B. Virag,
Acta Math. {\bf 194}, 1-35(2005).

\bibitem{simon10}
B.Simon, 
{\it Szego's Theorem and Its Descendants: Spectral Theory 
for L2 Perturbations of Orthogonal Polynomials},  
(Princeton University Press  2010).



\bibitem{Luzin98a}
Abe Shenitzer,N. Luzin,
"Function: Part I",
The American Mathematical Monthly, {\bf 105}, 59-67(1998).

\bibitem{Luzin98b}
N. Luzin,
"Function: Part II",
The American Mathematical Monthly, {\bf 105}, 263-270(Mar., 1998).


\bibitem{def}
We can also define the $Z-$transform  
by negative power $z^{-n}$.
Then the function $Z(z)$ is analytic in outer domain of $|z|=1$,
the poles corresponding to damping oscillations appear in 
the inside the unit circle  $|z|<1$.


\bibitem{gammel73}
J.L. Gammel and J. Nuttal,
J. Math. Anal. Appl. {\bf 43}, 694-696(1973).
\bibitem{yamada13} H.S.Yamada and K.S. Ikeda,
"Analyticity of Quantum States in One-Dimensional Tight-Binding Model",
preprint (2014).


\bibitem{kahane85}
J.-P. Kahane, 
{\it Some Random Series of Functions, 2nd edition}, Cambridge
Studies in Advanced Mathematics, 5, 
(Cambridge University Press, Cambridge, 1985).









\end{thebibliography}
\end{document}